# A Model and Tool for Community Engagement

Case Study: Community Engagement in the Bisotun World Heritage Site

Ahmad Nasrolahi
Tech4Culture Phd program
University of Turin
Turin, Italy
ahmadnasrolahi@gmail.com

Vito Messina
Historical Studies Dept.
University of Torino
Turin, Italy
vito.messina@unito.it

Cristina Gena
Computer Science Dept.
University of Torino
Turin, Italy
cristina.gena@unito.it

## ABSTRACT

Participating local community in cultural heritage management has always been a concern since the Venice Charter so far (1964). In addition, the Faro Convention (2005) shifted focus from cultural heritage values to the values of cultural heritage for society. In this case, there is required to design a simple method and tool in order to facilitate the interaction between local people and authority. This paper introduces a method and tool (iCommunity) that are now using in community-centered approach in the Bisotun World Heritage Site.

The main goal of this application is to find a simple and applicable method and tool for inclusion the local community in decision-making processes in cultural heritage management. The proposed model is designed based on the general public participation principles, on the one hand, and the users' need and requirement in the world heritage site, on the other hand.

## CCS CONCEPTS

• Human-centered computing → Empirical studies in HCI;
• Applied computing → Arts and humanities

## KEYWORDS

Local Community, Mobile Application, Cultural Heritage, iCommunity Application

## 1 Introduction

One of the most important features of people participation approach is to achieve the maximum engagement of stakeholders in all stages of management. Nowadays, the concept of community engagement in cultural heritage management is accepted for almost everyone. If we accept that community engagement is good for cultural heritage sites, the problem is how to let people participate? Not only communities are not very aware of their rights on the cultural heritage, but also cultural heritage authorities are unwilling to involve people in their decision-making process.

Suppose, in an ideal society of course, people know that the cultural heritage is their properties and they have right to manage their assets by themselves, and authorities understand that they are not omni-knowledge and omni-potent to make decision, there is still an unsolved problem; no one knows "how" local community should be engaged in a decision-making process by a people-centered approach. This "how" is referring to two main issues related to community engagement approach; lack of recognized method and determining an appropriate tool. This paper is proposing a practical model and a simple tool for including local community in decision-making process in the Bisotun World heritage Site.

The site of Bisotun is located along one of the main routes connecting the Iranian Plateau with the Mesopotamian plain, and is associated with the sacred Bisotun mountain. There is archaeological evidence of human settlement that date from prehistoric eras to the Islamic period, including remains dating to the Median, Achaemenid, Parthian and Sassanian times: the most significant period is that ranging from the 6th century BCE to the 6th century CE, however.

The management system of the site, which is the Bisotun World Heritage Research Base, is a part of the national government (Ministry of Cultural Heritage, Tourism and Handicraft), which is responsible for protection, conservation, education and rehabilitation of the area. The main important task of the Research Base is to ensure safeguarding and protecting the cultural heritage properties in its own landscape zone. Moreover, one of the properties, the relief and inscription of Darius the Great, has been designated in the World Heritage List in 2006 [1].

The landscape zone covers more than 35000 hectares, which includes 150 cultural heritage properties registered in the national heritage list. In this area, more than 60000 people are living with the cultural heritage sites, thus, they are daily engaging with these heritage site's issues.

Like other cultural heritage institutions, the management department of the Bisotun World Heritage Site is used the existing social media (Mostly Instagram and WhatsApp) for interaction with the local community and users. But it was found that these social media don't meet all the needs and requirements of public participation approach. Because of and basically, these social media were not designed for a people-centered activity in the first place. Moreover, it is impossible to modify the function of exciting social media based on the user's needs and requirements. If we would like to follow the public's participation approach in different levels, step by step from informing to empowering, the used application must be adaptable to change [2]. It means that by user's feedbacks, the application must be easily modified based on their new needs. The direct involvement of final users in the design process also enables a design that support user's preferential choices [3], namely the broad class of cases in which the user can chose among two or more options, none of which is correct or incorrect but one of which can be preferred to others.

In addition, Social media are typically made for access from all over the world, which is not that important in public participation process which is literally in local level. Although the Bisotun is a world heritage site with outstanding universal values, but it is required to involve local community living in the core, buffer and landscape zones in the system. It means that the proposed model must aim at local people in a local scale.

The main goal of this application is to find a simple method and tool for inclusion societies in decision-making processes in cultural heritage management. The idea is to encourage different stakeholders, such as local people living in or around the museums and cultural heritage sites, to take active roles in all stages of participation (from planning to making decision). Furthermore, this model and application will provide sufficient information and clear data for direct and indirect education of users by holding different workshops and test and evaluate the proposal with finals users [4]. Data shown in the application will also help people to understand the reasons behind the implementation of planned activities by taking part in comments and talking with experts or professionals. In addition, it also aims to make the decision-making process clearer and more transparent by presenting voting functions and showing all comments for users. Finally, the outcomes (which include analyzed data collected by feedback, voting, communication, etc.) will help to understand the real needs and interests of different stakeholders in the Bisotun World Heritage landscape zone. This method and application are adaptable and applicable for implementing in other cultural heritage sites and museums.

## 1.1 Principles

*1.1.1 Informing:* the "public's participation goal of informing is to provide the public with balanced and objective information to assist them in understanding the problem, alternatives, opportunities and/or solutions" [5]. It means that all sorts of relative information must be publicly published. Moreover, 'informing' in the spectrum of public's participation will make sense when it acts as a part of the whole processes which are informing, consulting, involving, collaborating and empowering. Since Arnstein's Ladder of Citizen Participation in 1969, informing has always considered as a prime stage in people's participation theory. At that time, even until the early of twenty-one century, the authorities had a power to "inform" people or not, as they wished, but after the emergence of information age, people have independently access to almost all data they need all around the world. How are people supposed to be informed while they already know, if they would like to, of course?

It is interesting to say that sometimes (or probably usually) informing deceives the authorities as well people in participation process, in this form that informing stage itself considers as a kind of people participation. It is again highlighted that 'informing, consulting, involving, collaborating and empowering' must implement as a system to achieve people's participation.

*1.1.2 Consulting:* although most often 'consultation' considers as a part of participation process, but there is a huge gap between consultation and participation. "Asking or being asked for information and advice" are the implicit concept of consultation meaning, while participation means having a part, collaboration and of shared ownership or responsibility which is totally different with the meaning of consultation. Moreover, participation displays various forms of 'communication' and 'involvement' that imply a strong mutual connection. That's why some experts consider consultation as a weak form of listening, which is on the opposite side of participation [6].

*1.1.3 Involving:* is the main hidden-principle of 'participation' which means 'have or include (something) as a necessary or integral part or result' and 'cause to participate in an activity or situation.' In fact, participation without involving (or engaging) is meaningless. The goal of involving is "to work directly with the public throughout the process to ensure that public concerns and aspirations are consistently understood and considered" [7].

*1.1.4 Collaborating:* which means to work jointly on an activity or project. The aim of collaborating is "to partner with the public in each aspect of the decision including the development of alternatives and the identification of the preferred solution" [8]. In this stage, the authority will look to people for advice and innovation in formulating solutions and incorporate people's advice and recommendations into the decisions to the maximum extent possible.

*1.1.5 Empowering:* is considered as a 'promised land' in people's participation approach, where all participation practitioners wish to get there. Literally meaning, empowering is "give (someone) the authority or power to do something." Moreover, it means to "make (someone) stronger and more confident, especially in controlling their life and claiming their rights." The goal of empowering stage in people's participation is "to place final decision-making in the hands of the public" [9].

## 2 iCommunity Model and Tool

The iCommunity model is a method for community engagement process in the Bisotun World Heritage Site by using a web-app application as a tool. For designing this model, several aspects have been considered in order to meet the public participation needs and requirements. These 'needs and requirements' should cover the spectrum of public participation which were considered as the model principles.

After recognizing the general needs and requirements, the iCommunity model has been prototyped as a mobile application in order to find its straights and weaknesses, as. Well as for better understanding the users (local community, NGOs and Management Department) ideas about the model. After several modification based on the user's needs, the application is going to assess by heuristic evaluation which is ongoing process now (Figure 1).

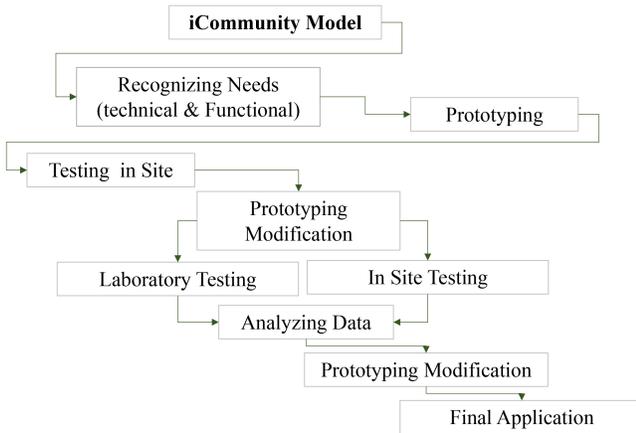

**Figure 1: Different Stages of Designing the iCommunity model/application**

For the informing function, iCommunity application is able to publish new and future activities in an appropriate way to let the people know about what is happening in the Bisotun World Heritage Site. Along with 'what is happening,' the complementary information such as the location, the ideas behind the activities, budget assessment, relative researches, etc. are attached to the posted activity.

Based on the published activities, the application must provide relative workshops and training course for improving local people's knowledge. Before the pandemic, every year, the Bisotun World Heritage Research Base held a number of specialized and general workshops for different age groups as one of their organizational duties. But during quarantine, they were unable to continue on that way. The proposed smartphone application should be able to organize these workshops and events in the form of 'in site' or 'online.' In site events will post in the main page to inform people for participation due time and online workshops are published on the application. In the latest one, the users have access to downloadable documents. Most often, other cultural heritage institutions hold workshops and events which are also useful for the Bisotun World heritage Site. Through the application, iCommunity's admin shares the link to allow the users for participation. For the consulting purpose, these functions are considered: comment, message, talk to expert and asking for permissions. Each posted activity has a space for the user's comment. In this section, the users are able to post the for and against ideas on the projects and activities. Like Instagram, other users can read and participate in the topics raised in the comments. For more connection, the users also can receive and send direct text to each other via the message function [10].

Furthermore, every user can consult with an expert in the case of needing more information and discussion. Basically, in the Bisotun World Heritage Research Base, there are different department that are working on different topics and every activity and project has a specific expert who is in charge of the given project. The users must have direct access to the project manager. Since these project managers properly know all the activity information, they are the best one for asking and arguing.

Users are able to upload their documents in the form of image, video, voice and pdf files in the add user's experience section. Local people often have valuable information about the cultural heritage site that is beneficent for conservation and management of the given heritage site. On the iCommunity, users can share old pictures, stories, legends and written documents to experts for using them in the projects and activities. This function aims at involving the community in the process.

Every year in some special events, the Bisotun World Heritage Site needs temporary recruitment without payment in different position. For example, in Nowruz holidays, which the number of visitors is increasing, the Site needs more tourist guides. Thus, the positions publish in the voluntary activity section in order to ask enthusiasts and local people to involve. These voluntary positions are not limited to Nowruz and tourist guides, the World Heritage Site always requires various specialists including archaeologist, researchers, students, carpenters, etc. (Figure 2).

| Purpose | Functions | Specifications |
|---|---|---|
| Access to the App | Personal Account | Login via google, Instagram, Facebook, Twitter or directly |
| | Institutional Account | |
| Informing | Location | Mobile GPS, google map |
| | Workshops | Upload document, image, video and share link |
| | New Activities | Upload document, image, video and share link |
| | Message | Sending text and notification |
| | Data Analysis | Access to Outcomes |
| | More Information | Access to Documents Database |
| Consulting | Message | Private connection to other users |
| | Talk to Expert | Private Connection to Admin |
| | Asking for Permission | Online Forms for Submitting Requests |
| Involving | Comment | To engage in new activity |
| | User's Experience | Adding user's data on a given activity |
| | Monitoring | Helping the authority in monitoring the site |
| | Talk to Expert | Private Connection to Admin |
| | Voluntary Activity | Taking part of an activity |
| Collaborating | Voting | Active participation in decision-making |
| | Monitoring | Helping the authority in monitoring the site |
| | Voluntary Activity | Taking part of an activity |
| Empowering | Voting | Active participation in decision-making |
| | Data Analysis | Access to Outcomes |
| | Propose Activity | Listening to user's ideas |

**Figure 2: Proposing Various Functions Based on People Participation Principles**

Those all functions aim at the collaborating and empowering propose. Listening to user's voice, gathering different ideas, arguing, voting and publishing the outcomes in the data analysis function lead us to engage the local community in the decision-making processes. The iCommunity application at least will provide an appropriate condition for achieving the community empowering purpose (Figure 3).

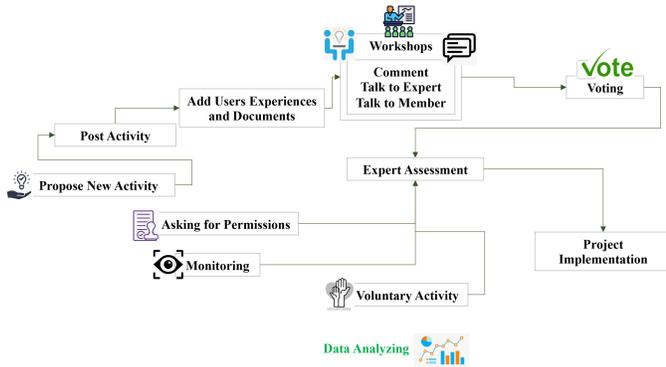

**Figure 3: Proposed Model for Community Engagement in the Bisotun World Heritage Site**

## 3  Conclusion and Ongoing Works

iCommunity is an application for facilitating interactions among different users involved in the Bisotun World Heritage activities. It helps the site's authority to include effectively the local community in decision-making processes by a people-centered approach. This project is part of a PhD study aimed at using new technology for public participation in cultural heritage management.

We are still working on iCommunity method and tool in the Bisotun World Heritage Site to fix bugs and modify it. In order to measure various aspects of the application, a heuristic evaluation has been done by master students of software engineering in the computer department of University of Turin that will apply after analyzing. In addition, the usability test will be done by summer in the Bisotun World Heritage Site.

### ACKNOWLEDGMENTS

This project has received funding from the European Union's Horizon 2020 research and innovation programme under the Marie Skłodowska-Curie grant agreement No 754511 in the frame of the PhD Program Technologies for Cultural Heritage (T4C) held by the University of Torino. Ahmad Nasrolahi would like to express his sincere gratitude to his research supervisors for their guidance, patience and encouragement.